# *A General Framework for Disturbance Compensation in Airborne and Seaborne Gravimetry Supported by Machine Learning*

**Authors:** Lorenzo Iafolla[*1], Massimo Chiappini[1], Roberto Carluccio[1], Stefano Chiappini[1], Emiliano Fiorenza[2], Carlo Lefevre[2], Pasqualino Loffredo[2], Marco Lucente[2], Francesco Santoli[2]

[1] Istituto Nazionale di Geofisica e Vulcanologia (INGV), Sezione Roma 2, Via di Vigna Murata 605, 00143, Rome, Italy

[2] Istituto di Astrofisica e Planetologia Spaziali (IAPS), Istituto Nazionale di Astrofisica (INAF), Via del Fosso del Cavaliere 100, 00133, Roma, Italy

# Abstract

High-accuracy measurement systems operating in complex operational environments, such as airborne and seaborne gravimetry, are severely affected by interacting external influences (disturbances) including temperature variations, inertial accelerations, and platform rotations. This work aims to develop and prove a general method for compensating such disturbances beyond the limits of conventional approaches.

We present a general framework based on three pillars: a multi-sensor system, supervised machine learning, and a dedicated laboratory -training platform-. This measurement technique was investigated through two experimental case studies relevant to airborne and seaborne gravimetry: (i) temperature and thermal-gradient rejection in a high-sensitivity tri-axial accelerometer, and (ii) pitch and roll rejection for vertical acceleration estimation.

The experiments highlighted that, although the general framework provides a useful guideline, its application to airborne and seaborne gravimetry is challenging and not straightforward, requiring the development of dedicated and innovative experimental solutions. This difficulty arises from the specific nature of the measurand—gravity—which cannot be easily varied under controlled conditions. In this work, we present a dedicated approach to addressing these challenges.

[*] Corresponding author: lorenzo.iafolla@ingv.it; lorenzo.iafolla@outlook.it

# 1 Introduction

Regardless of the quantity being measured, a primary obstacle to achieving high-accuracy measurements are disturbances—also referred to as “modifying inputs” or “interfering inputs”—. These are external variables (e.g., temperature, pressure, humidity) that introduce unwanted variations or errors, thereby compromising both precision and reliability of measurement systems (Hansman, 1999).

The principles and techniques presented here are broadly applicable to a wide range of measurement contexts and disturbances. For clarity, however, we illustrate our discussion with a specific application: gravimetry on moving platforms (mobile gravimetry), such as airborne or seaborne gravimetry. Gravity measurements play a crucial role in geophysical exploration, navigation, and Earth system science, providing insights into subsurface density variations and mass distribution (Philip Kearey et al., 2013). Mobile gravimetry, conducted from airborne (Forsberg and Olesen, 2010; Jensen and Forsberg, 2018; Luo et al., 2022) or marine (Ai et al., 2023; Shinohara et al., 2018; Vu et al., 2024) platforms, offers the significant advantage of accessing remote or otherwise inaccessible regions, enabling efficient data collection over large and diverse areas (Dadrass Javan et al., 2025). We focused specifically on gravimetry onboard moving platforms because it exemplifies a measurement technique that demands extremely high accuracy in the presence of significant disturbances. Depending on the application, required accuracies in gravimetry range from a few milligals ($10^{-6}$ g ~ $10^{-5}$ m/s$^2$) to less than a microgal ($10^{-9}$ g ~ $10^{-8}$ m/s$^2$) (Niebauer, 2015). Even on the ground, under stable environmental conditions, the omnipresent micro-seismic signal generated by ocean waves, often exceeding amplitudes of $10^{-5}$ m/s$^2$ (Lederer, 2009; Van Camp et al., 2017), can overcome these accuracy thresholds. On moving platforms, the problem intensifies: motion-induced noise and environmental variability can exceed the gravity signal by several orders of magnitude, making high-precision measurements extremely challenging (Schwarz and Li, 1997; Wei and Schwarz, 1998). Additional challenges arise because moving platforms are non-inertial reference frames, where apparent (inertial) acceleration signals overlap with gravity signals. Moreover, environmental factors such as temperature, air pressure, and humidity often fluctuate during operation, further degrading measurement quality. The phenomenon is illustrated with the two schemes in Figure 1. As a consequence, state of the art precision in airborne gravimetry is about one milligal (Dadrass Javan et al., 2025), whereas gravimeters could achieve sub-microgal precision under disturbance-free conditions. This large gap illustrates the metrological challenge addressed in this work: compensating for disturbances that prevent high-performance sensors from operating near their intrinsic accuracy in moving-platform applications.

This paper aims to advance methods for correcting disturbances in measurement systems for gravimetry and accelerometry. To this end, we present and discuss a machine learning–based approach for calculating output corrections and enabling high-accuracy measurements in complex systems. The specific nature of the measurand—gravity—required the development of dedicated and innovative experimental solutions, as generating appreciable variations in gravity under controlled conditions is inherently difficult. In addition, initial experimental implementations of this approach for airborne strapdown gravimeters are presented.

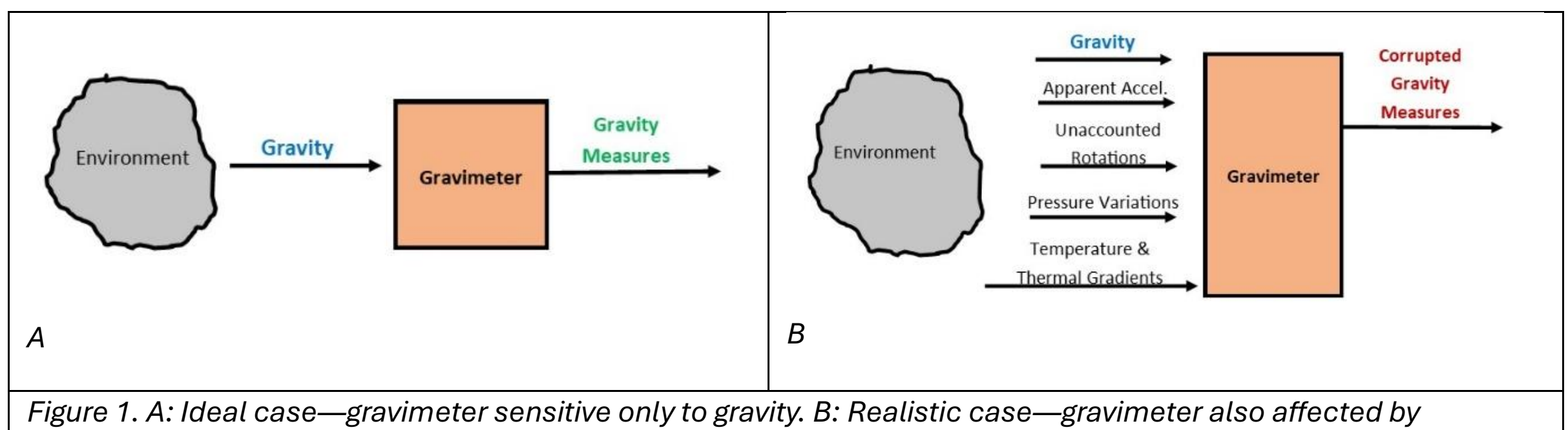

*Figure 1. A: Ideal case—gravimeter sensitive only to gravity. B: Realistic case—gravimeter also affected by disturbances.*

# 2 Background

Several general methods exist for correcting disturbances (Doebelin, 2004; Morris and Langari, 2021a). Here, we focus on few approaches related to the design and implementation of a measurement system, leaving aside data-processing techniques at this stage.

One method is -**inherent insensitivity**-, which means that the sensor is designed to respond only to the desired signal (e.g., the gravity signal). For instance, because temperature can interfere with the functioning of a gravimeter (Belwanshi et al., 2022; Bourgeteau-Verlhac et al., 2017; Fores et al., 2017; Weerasinghe and Prasanna, 2021), it should be constructed using designs and materials whose properties are minimally affected by temperature (Tariq et al., 2024; Xiao et al., 2016). However, this method is sometimes inapplicable or only partially effective. For example, as a direct consequence of the Einstein Equivalence Principle, it cannot be used to make a gravimeter selectively sensitive to gravity while rejecting inertial accelerations.

-**Opposing input**- and -**physical insulation**- are two other conventional methods, involving either physical compensation for disturbances or insulation of the measuring system from them. For example, temperature variations can be mitigated by using an active, closed-loop thermal control system or by insulating the sensor, such as by mounting it inside a vacuum flask (Yang et al., 2022). Another important disturbance in -moving platform- gravimetry arises from unaccounted platform rotations, which alter the alignment between the gravity vector and the gravimeter's sensitive axis. In the opposing input method, a stabilizing gimbal suspension can be employed to maintain constant alignment with the vertical (Glennie et al., 2000). However, both opposing input and physical insulation increase system complexity, often adding volume, weight, and power consumption. This makes them unsuitable for applications where such resources are limited, as in seaborne and airborne gravimetry.

Another method is -**calculated output corrections**-, in which disturbances are analytically accounted for by modeling the sensor's response to them. In this approach—often employed in "smart sensors" (Morris and Langari, 2021b) and commonly referred to as "sensor fusion" (Liggins II et al., 2009)—disturbances are measured using dedicated secondary sensors, and their effects are removed during data processing (see Figure 2). This typically requires a calibration phase, during which the analytical relationship (we will refer to it as "system model") between the sensor output and the disturbance is determined with a certain degree of approximation (Morris and Langari, 2021c). For instance, in the case of thermal variations, a linear relationship between the gravity sensor output and its temperature can be established during calibration, allowing the effect to be calculated based on thermometer readings (Guo et al., 2023; Han et al., 2020; Huang et al., 2023). However, the same approach becomes far less straightforward for complex combinations of disturbances—such as platform rotations, inertial accelerations, and variations in thermal gradient, pressure, and humidity. In strapdown airborne gravimetry, for example, a relatively complex procedure based on a Kalman filter (Welch and Bishop, 1995) is used to fuse data from accelerometers and gyroscopes—i.e., a strapdown Inertial Measurement Unit (Titterton and Weston, 2004)— with data from GNSS receivers (Jensen, 2018). A major limitation of this method is that the mathematical model required for calculating corrections becomes increasingly complex as the number of interfering disturbances and secondary sensors increases. Our method aims to overcome these limitations by leveraging machine learning to construct data-driven models of measurement systems that can capture highly complex disturbance interactions and enable more effective compensation.

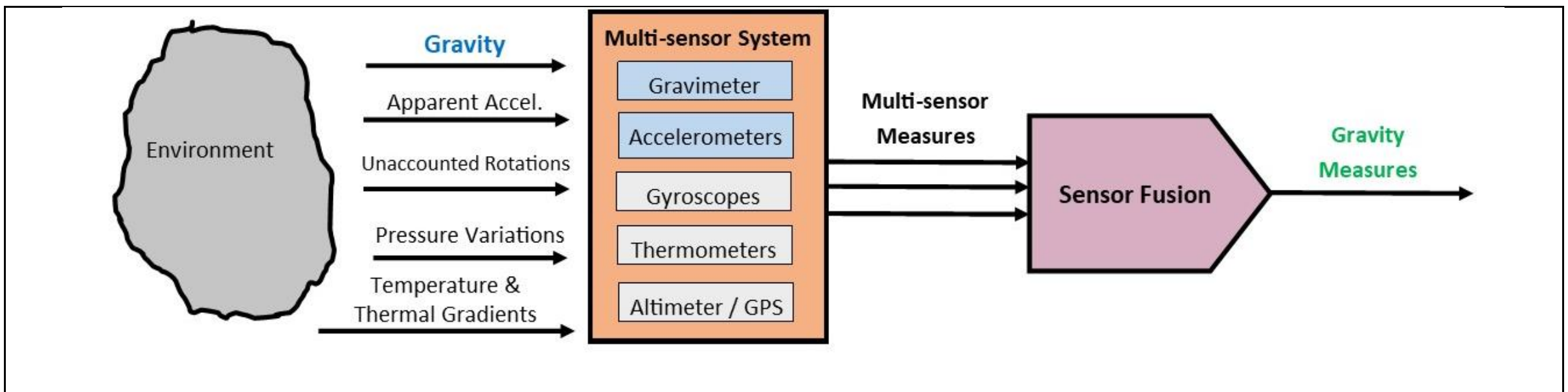

*Figure 2. Within the calculated output correction method, a multi-sensor system measures both the measurand (gravity) and relevant disturbances (apparent accelerations, unaccounted reference-frame rotations, etc.). A sensor fusion algorithm then reconstructs the gravity measurements from the multi-sensor outputs.*

# 3 Experimental Applications

To investigate the principles and advance the understanding of the proposed framework and its application to gravimetry and accelerometry, we considered two experimental case studies. The first concerns the rejection of temperature and thermal-gradient effects in the output of a tri-axial accelerometer. Although this experiment was previously described and validated in detail (Iafolla et al., 2023, 2024c), it is revisited here from the perspective of the general framework proposed in this paper, with emphasis on its role as an example of a simplified training strategy for disturbance-induced error reconstruction. The second is an original case study focusing on compensating pitch and roll rotations in vertical acceleration measurements (Iafolla et al., 2024b, 2024a). Further applications of the method, not discussed in this work, concerned spaceborne accelerometers (Fusco et al., 2024).

## 3.1 Temperature rejection

Temperature and thermal gradients variations affect the output of a tri-axial accelerometer by altering the properties and geometry of its mechanical components.

The dominant effect arises from temperature-induced changes in the elastic properties of the sensing elements. This influence is typically stronger in gravimeters—that is, in the accelerometer component primarily aligned with gravity—than in the horizontal components. To correct this specific disturbance, it is sufficient to monitor the temperature of the sensing element of the accelerometer, measured with a dedicated thermometer.

A secondary effect is caused by thermal gradients that deform the accelerometer's mechanical frame. For example, if one side of the frame expands more than the opposite, one sensitive axis might tilt slightly with respect to the gravity vector, thereby producing a change in the accelerometer output. In this case, multiple thermometers are required to capture the gradient and to compute the appropriate correction.

Deriving an analytical model for these corrections is far from straightforward, and residual imperfections in sensor behavior are difficult to capture analytically. Our method can address both issues by using machine learning to model the effects of these disturbances directly from the data and integrate them into the correction process.

## 3.2 Pitch and roll rejection for vertical acceleration measurement

In airborne gravimetry, correcting unaccounted rotations of the reference frame is a fundamental step because such rotations alter the relation between the accelerometer sensitive axes and the gravity vector. These rotations must be measured using gyroscopes, which typically provide angular rate measurements rather than orientation angles, making the processing more involved. Analytical models describing how to combine accelerometer and gyroscope data are well established (Titterton and Weston, 2004) and represent a natural starting point. However, relying exclusively on these models does not account for hardware imperfections, misalignments, or other sources of residual error. The presented method can complement analytical models by incorporating such characteristics into the data-processing stage through machine learning.

# 4 Methods and Materials

## 4.1 Data-Driven Framework for Disturbance Compensation

The presented method is based on three pillars: 1) Multi-sensor system, 2) Machine Learning, and 3) Training platform.

As in -calculated output corrections- method (see Section 2), the idea behind the first pillar is to simultaneously measure both measurand (i.e., quantity being measured) and disturbances using multiple sensors of different types (Liggins II et al., 2009), see Figure 2. Thus, we can distinguish primary sensors (highlighted in light blue in the scheme of Figure 2), which are mainly sensitive to the measurand, and secondary sensors (in gray), which are mainly sensitive to disturbances. In the instance of airborne gravimetry, the primary sensors are accelerometers, in which the component primarily aligned with gravity acts as the gravimeter. As explained before, these sensors are sensitive not only to gravity variations but also to disturbances. On the other hand, the secondary sensors are not sensitive to gravity, but they are to disturbances; for instance, they might be thermometers, barometers, gyroscopes, etc. The combined output of the multi-sensor system consists of multiple signals (i.e., the measures from each sensor) which are correlated one with the other in an intricate way. The consequent challenge is to recover the measurand signal (the gravity, in the instance of airborne gravimetry) from such complex output.

The second pillar addresses this challenge using Machine Learning (Géron, 2019), a type of artificial intelligence built on algorithms that can be trained on data—referred to as “training data”. Once the Machine Learning algorithm is trained and tested, the multi-sensor system can be deployed in the operational context and its output data can be processed to retrieve the desired signal. However, the need for training data remains unaddressed.

The third pillar, the training platform, addresses this need. The training platform is a laboratory settings able to reproduce the operational context. For example, if the multi-sensor system is supposed to be used onboard a boat, the training platform should physically replicate the oscillation movements of the deck of the boat.

In practice, the presented method is implemented in three phases as shown in Figure 3. During the first phase, the multi-sensor system is temporarily installed on the training platform to gather the necessary training data. Once enough data have been collected, the process moves to the second phase, during which the ML algorithm is trained yielding a ML model. Finally, during the third phase, the multi-sensor system is deployed in the operational context—such as onboard a boat or an airplane—and the data collected are processed using the previously-trained ML algorithm.

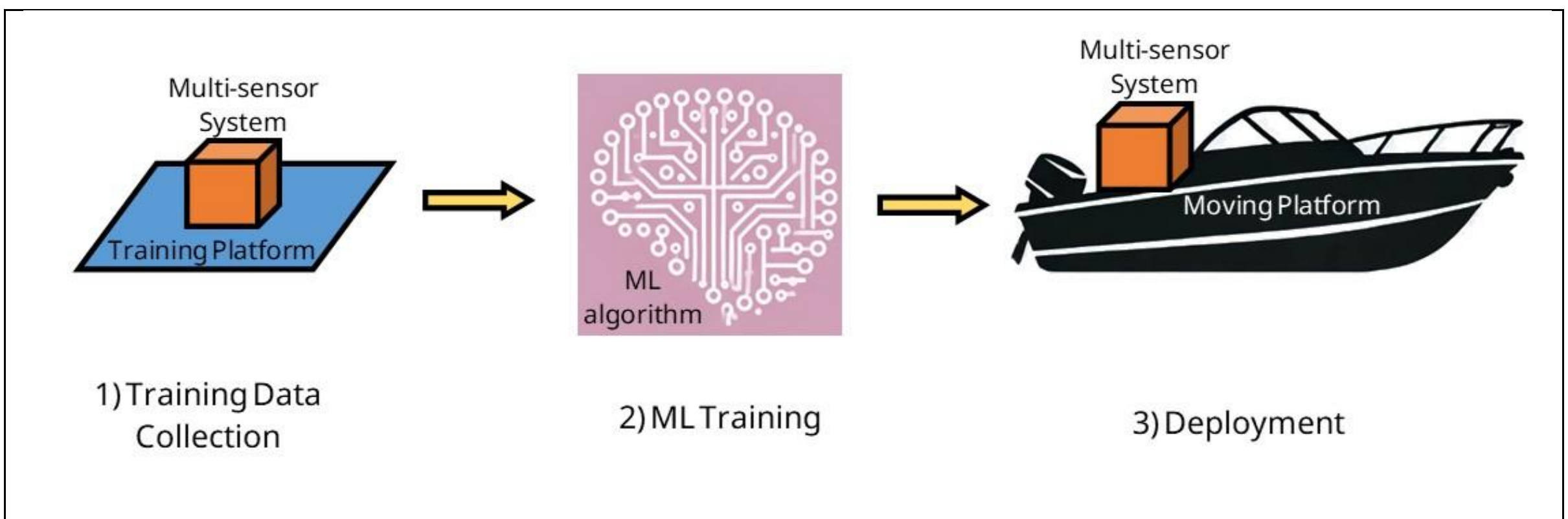

*Figure 3. The method consists of three phases: 1) Training data collection; 2) ML training; 3) Deployment in the operational context.*

### 4.1.1 Multi-sensor System

The multi-sensor system must guarantee that all signals (i.e., disturbances and gravity signal) required to reconstruct the measurand (i.e., the gravity signal) are measured during operation. This is because each sensor provides only part of the information, much like a piece of a puzzle. Thus, besides the primary sensors, at least one secondary sensor is required for each disturbance with a non-negligible impact on the measurement. For this reason, designing a multi-sensor system requires a thorough understanding of all environmental disturbances and of their effects on the output of the primary sensors. This understanding will also be crucial to assess, by computing error propagation, the required accuracy of each sensor.

Another important constraint arises from the need to fuse signals from all sensors. This is because each sensor generates a time series describing the evolution of its associated variable. Consequently, proper fusion requires that these time series are accurately synchronized; otherwise, measurements from different instants would be wrongly combined. To prevent this, the acquisition architecture must operate in real-time, with delays constrained and characterized to levels that do not compromise data fusion. As a practical guideline, synchronization errors should be kept substantially below the shortest sampling period among the acquired signals, for example within approximately 1/10 of it.

In the example of airborne gravimetry, the main disturbances are unaccounted frame rotations, inertial accelerations, and temperature variations. Other error sources, such as synchronization errors or sensor noise, are not considered “disturbances” as defined in Section 1. The primary sensor is typically a three-axis accelerometer, although a gravimeter—essentially a single-axis accelerometer—in principle could suffice. To compensate for the main disturbances listed above, the minimum set of secondary sensors includes a three-axis gyroscope, a GNSS receiver, and thermometers, while additional sensors (e.g., pressure or humidity) can further enhance compensation. Indirect effects must also be considered; for instance, temperature gradients can deform the physical structure of the system, thereby altering the relative orientation of the accelerometer and gyroscope axes. Addressing these effects requires a detailed characterization of the temperature field, achieved with multiple thermometers or thermal cameras (Iafolla et al., 2024c).

### 4.1.2 Machine Learning

Machine learning (ML) is a set of methods for deriving a mathematical model from data (training data), which can then be used to make predictions on new inputs. ML methods are typically divided into *classification*, where the output is a discrete class, and *regression*, where the output is a continuous variable. Since the presented framework involves producing continuous measurements, only regression algorithms are considered. A further distinction is made between supervised, unsupervised, and reinforcement learning. Here we focus exclusively on supervised ML and refer the reader to (Géron, 2019) for an overview of the other approaches.

A supervised ML project usually consists of three steps: training, testing, and deployment. During training, the algorithm adjusts its internal parameters to minimize the discrepancy between predicted output and the desired output (label), which is provided in the training dataset. The resulting ML model can then generalize to new, unseen inputs. In the testing phase, its predictive accuracy is assessed on a separate labeled dataset. In the airborne gravimetry example, the desired output would be the gravity signal, which must be available for both training and testing. Finally, in deployment, the trained ML model is applied to new data for which the true outputs are unknown.

There is no single machine learning algorithm capable of addressing all possible problems, this is formalized in the No Free Lunch theorem (Wolpert and Macready, 1997). Consequently, the challenge is to identify the algorithm that best fits the characteristics and requirements of the specific task at hand. The selection process can therefore proceed step by step, progressively narrowing the choice by considering families of algorithms whose properties align with the problem's features. Here, we can focus on supervised regression algorithms which can be broadly grouped into two categories: conventional ML algorithms and Artificial Neural Networks (ANNs).

-**Conventional ML algorithms**- include methods such as Linear Regression, Support Vector Machines (for regression), Gaussian Processes, and Decision Trees. These approaches heavily rely on feature engineering, i.e., selecting and transforming relevant input variables to improve predictive performance. Effective feature engineering typically requires a deep understanding of the physical process that generated the data, so that meaningful features can be identified and irrelevant or redundant ones discarded. For example, in airborne gravimetry, it may be useful to transform the angular rates measured by the gyroscope into rotation angles. The main advantages of conventional ML algorithms are relatively fast training, modest computational requirements, and easier interpretability. However, they often struggle with highly complex tasks.

-**Artificial Neural Networks**- are inspired by biological neural systems. Within this category, deep learning refers to architecture with multiple hidden layers of artificial neurons. Deep learning methods can handle minimally preprocessed data, effectively performing feature extraction automatically. They excel in complex tasks such as image classification or natural language processing, but they require careful hyperparameter tuning, high-performance computing (usually with GPUs), and long training times. Nevertheless, their internal decision-making process is difficult to interpret, which is why they are often referred to as "black boxes."

In the proposed framework, the choice between ML architectures is not prescribed a priori; rather it is treated as part of the application-specific model-selection process. In practice, ML implementation can rely on in-house code developed in a general-purpose programming language, together with standard open-source scientific libraries; this was the approach adopted in the present work, where the processing was performed offline using Python libraries such as NumPy, SciPy, Pandas, and Scikit-learn, Keras, etc.

### 4.1.3 Training Platform

The third pillar of the method is the training platform, which provides the labeled data required for supervised machine learning. The platform has actuators capable of physically varying both the disturbances and the measurand. In this way, it reproduces the operational environment in a controlled laboratory setting while ensuring that the full range of expected conditions is reproduced.

Similarly to the design of the multi-sensor system, detailed knowledge of the target application and its relevant disturbances is essential for developing the training platform. Ideally, its design should mirror that of the multi-sensor system: for each sensor (primary and secondary), a corresponding actuator should be available to induce variations in the associated physical quantity—for example, thermal systems to impose temperature and thermal gradients changes for thermometers, or mechanical actuators to rotate the system for gyroscopes.

Most ML algorithms exhibit poor generalization performance when dealing with extrapolation, that is, when predictions are required outside the domain spanned by the training data. To mitigate this issue, the training platform must ensure that both the measurand and all disturbances vary across a sufficiently broad domain, so that the collected training dataset encompasses the full range of expected operating conditions. This strategy mitigates the risk of extrapolation during real-world deployment and enables more robust learning. For example, if the expected operational temperature interval is known, the training platform should reproduce at least that interval, so that the ML algorithm can learn within the same domain in which it will later be applied.

The measurand itself must also vary during training and, crucially, be independently measured by a high-accuracy reference instrument integrated into the platform ("reference sensor"). This reference measurement provides the "desired output" required to construct labeled datasets for both training and testing. The reference sensor must be acquired in real time and strictly synchronized with the multi-sensor system to guarantee proper data fusion.

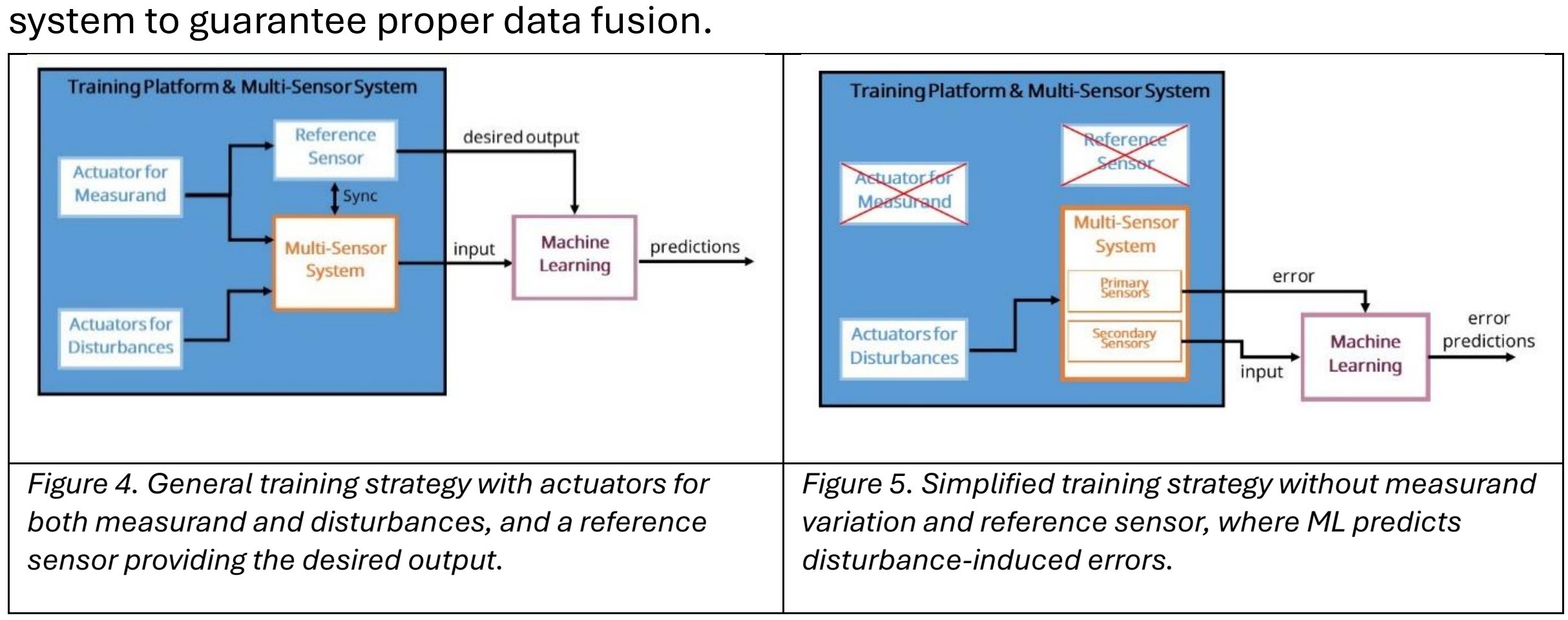

*Figure 4. General training strategy with actuators for both measurand and disturbances, and a reference sensor providing the desired output.*

*Figure 5. Simplified training strategy without measurand variation and reference sensor, where ML predicts disturbance-induced errors.*

Two different strategies can be adopted for training data acquisition, as illustrated in Figure 4 and Figure 5. Each strategy shapes the design of the training platform and defines how the ML algorithm is trained.

The first strategy (Figure 4) is the most general and has already been outlined in the previous sections. In this case, both the measurand and the disturbances are varied by actuators. The measurand is simultaneously measured by a high-accuracy reference sensor, which provides the desired output for supervised training. The ML algorithm is therefore trained to predict the measurand directly from the input signals of the multi-sensor system.

The second strategy (Figure 5) is a simplified variant that relies on the assumption of linear superposition of errors. In this case, no actuator is used to vary the measurand and therefore no reference sensor is needed. Disturbances are still induced, causing variations in the outputs of both primary and secondary sensors. The ML algorithm is trained by feeding it with the secondary sensor readings whereas the desired outputs are the disturbance-induced deviations of the primary sensor readings. In other words, the algorithm learns to predict the error caused by disturbances rather than the measurand itself. Real measurand variations, for example those caused by seismic events, environmental vibrations, or unmeasured disturbances, must therefore be minimized, filtered, or otherwise kept negligible with respect to the induced disturbance effects; otherwise, they would contaminate the training labels. During operation, the corrected measurand can then be calculated by subtracting the predicted error from the primary sensor output. This strategy is not applicable when disturbances interact with the primary sensors output in nonlinear ways.

## 4.2 Experimental Setups

In this section, we briefly describe the two experimental setups considered in Section 3. The temperature-rejection setup and dataset were preliminary reported in previous work (Iafolla et al., 2024c, 2023). The pitch/roll rejection setup is presented here as an original case study; the corresponding dataset is publicly available (Iafolla et al., 2024a).

### 4.2.1 Temperature rejection in static configuration accelerometers

We made the **Multi-sensor System** by modifying a high-accuracy tri-axial accelerometer designed and built as an early prototype of the space-borne Italian Spring Accelerometer (ISA) (Iafolla et al., 2010; Santoli et al., 2020). Background noise spectral density of each sensitive element was about 10 $\mu$Gal/√Hz ($10^{-8}$ g/√Hz ~$10^{-7}$ m/s$^2$/√Hz) within the operational bandwidth [ $5\times10^{-5}$ Hz to 10 Hz]. The major upgrade consisted of integrating 11 high-precision (better than $10^{-3}$ °C with pair-wise Pearson correlations of their measures better than 0.9998) thermometers strategically displaced in the accelerometer enclosure. The criteria for deciding the thermometers positions was to capture all temperature-field variations (including thermal gradient) that could have affected the acceleration measurement. We made the thermometers using platinum PT10000 resistors whose changes were measured through a Wheatstone Bridge supplied by a reference voltage (Analog Devices, REF195). The following electronic stages were a high-precision analog-to-digital converter (Texas Instruments ADS1263) and a microcontroller (Espressif, ESP32-Wroom-32). The latter was programmed to time-stamping data from both the thermometers and the accelerometer and send them to a desktop computer for data storage.

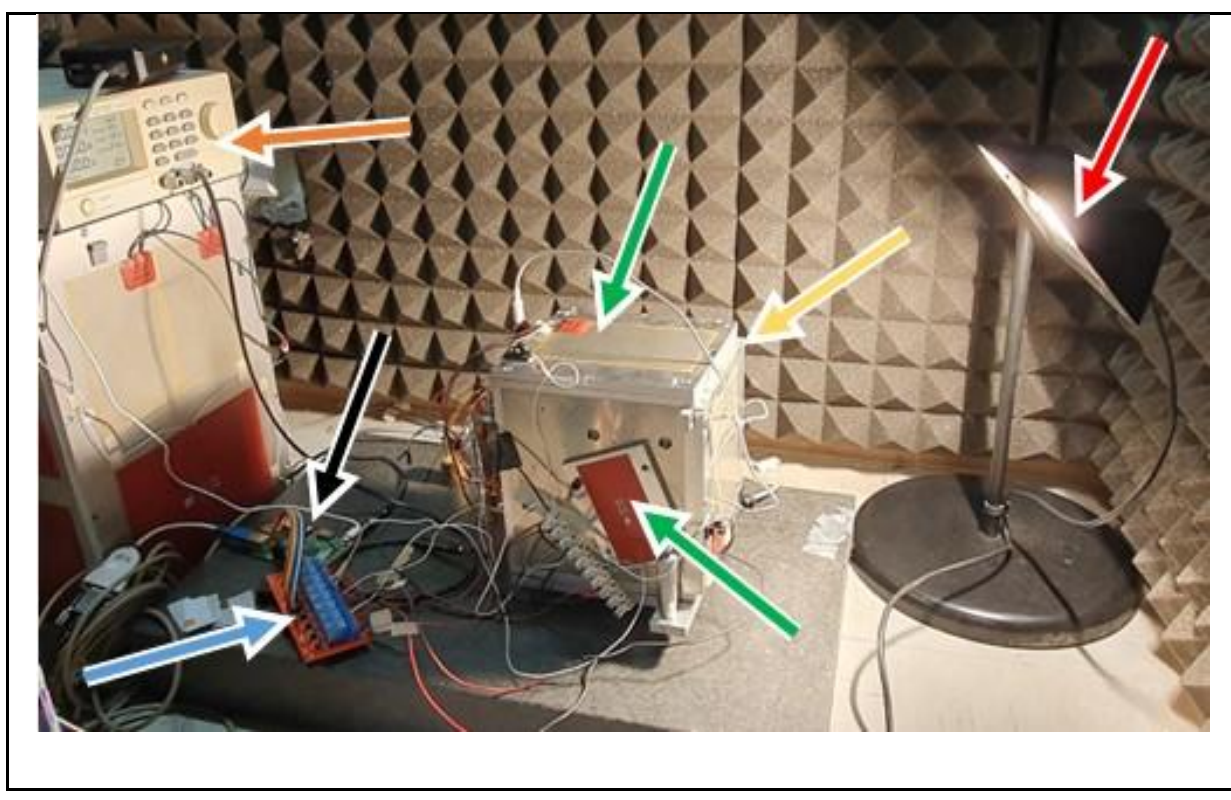

*Figure 6. Experimental setup for temperature rejection. The accelerometer is indicated by the yellow arrow; two of the six heating mats by green arrows; the lamp by a red arrow; the relays by blue arrows; the power supply by an orange arrow; and the Raspberry Pi (barely visible) by a black arrow.*

**Machine Learning**. We first synchronized the raw data from different sensors by interpolating temperatures to the accelerometer sampling rate and aligning all records on a common timeline. Afterward, we low-pass filtered the data, down-sampled them to 1 Hz, and dropped out spurious events such as earthquakes or anthropic activities. We calibrated all sensor outputs to standard units and normalized them to comparable scales. To provide richer information to the ML algorithm, we prepared the inputs using three feature engineering strategies: 1) instantaneous values, 2) time derivatives, and 3) short temperature sequences arranged into matrices, enabling the models to capture both current readings and temporal dynamics.

We trained different regression algorithms to reconstruct the error (as explained in Section 4.1.3, Figure 5). Linear regression was implemented in both simple and multivariate form, while feed-forward neural networks (FFNN) were used to capture more complex relationships. Each accelerometer axis was modeled independently, resulting in one machine learning model per each of them.

We designed the **Training Platform** in the simplified form as described in Figure 5 of Section 4.1.3. Thus, the accelerometer operated in a static configuration and its output mainly changed because of disturbances, i.e., temperature and thermal gradients variations. These were induced by two different types of actuators: 1) heating mats attached to the faces of the accelerometer enclosure, 2) heating lamp moved around to heat the device from different directions. The heating mats were randomly switched on and off to generate random temperature and thermal gradient fluctuations. To achieve this, we powered the mats, which are electrical resistors, through Raspberry Pi-controlled relays and a bench power supply (see Figure 6). By programming the Raspberry Pi to automatically control the heating mats, temperature and thermal-gradient variations could be generated over extended periods without operator intervention. This automation allowed the full data acquisition to last for more than two months. The thermal training platform generated temperature and thermal-gradient variations around room temperature, approximately 20 °C. The accelerometer sensing-elements showed peak-to-peak temperature variations of approximately 2 °C, whereas the accelerometer enclosure showed peak-to-peak variations of approximately 6 °C. The maximum temperature difference between thermometer pairs was approximately 4 °C. No active cooling was used; after heating phases, the system cooled passively toward room temperature. Because the heating mats were randomly switched on and off to generate non-periodic thermal disturbances, no fixed temperature step size was imposed.

### 4.2.2 Pitch and roll rejection for vertical acceleration measurement

We built the **Multi-Sensor System** upon that described in Section 4.2.1, further comprising a commercial tri-axial Fiber Optic Gyroscope (FOG, Emcore TAC-450) acquired by an ESP32 board at 10 Hz (see Figure 7). To guarantee synchronization between the acquired signals, the master ESP32 board, that acquiring the accelerometer, broadcasted an electronic Sync pulse to the slaves ESP32 boards, i.e., those acquiring the Gyroscope and reference sensors. Both master and slave boards recorded their own timestamps at the moment of transmitting (for the master) or receiving (for the slaves) the Sync pulse, this enabled data synchronization in post processing with accuracy better than 1 ms.

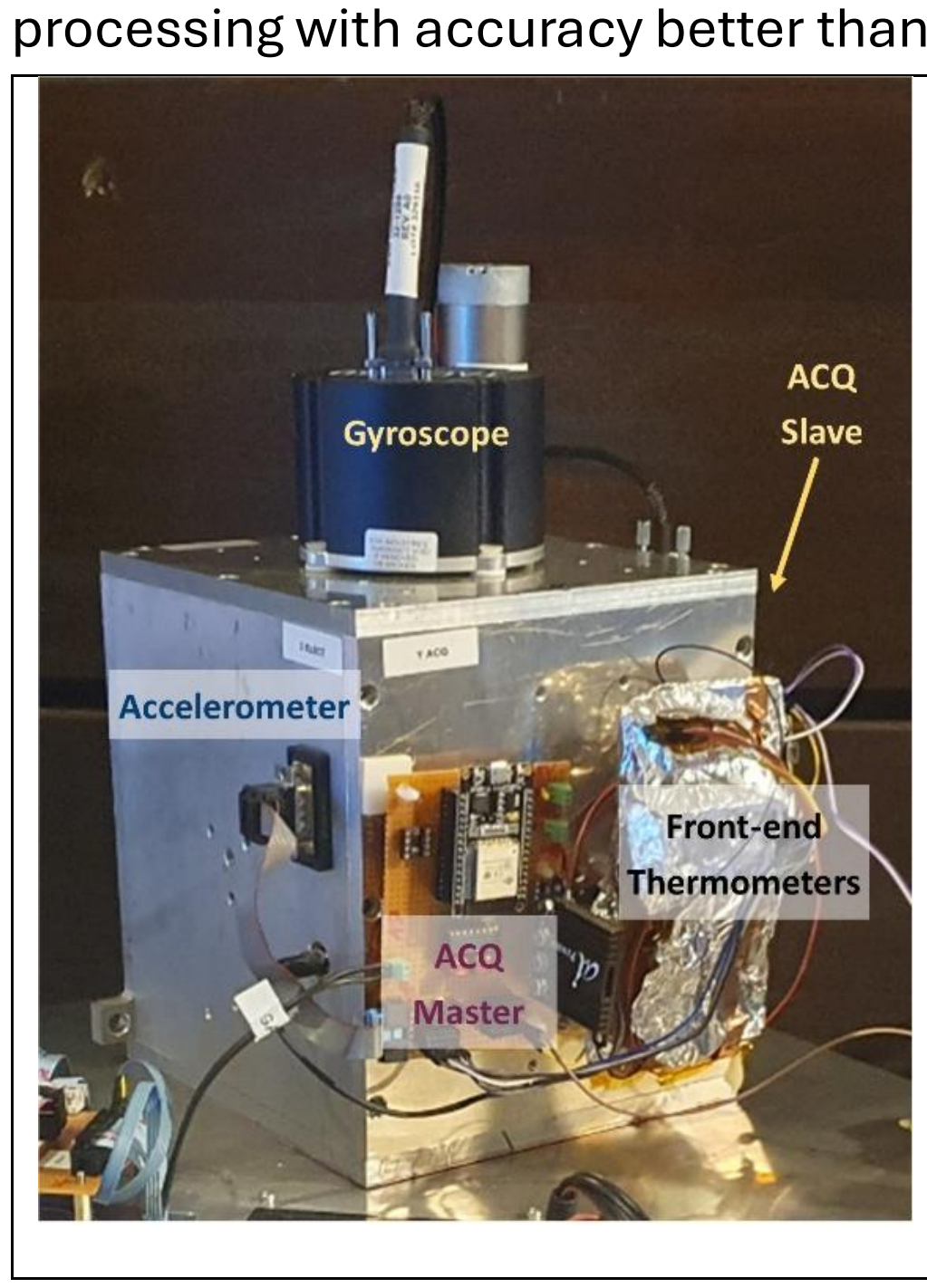

*Figure 7. Multi-sensor system used for the - Pitch and roll rejection for vertical acceleration measurement – experimental setup. The main components of the system, including the front-end and acquisition electronics, are indicated by captions.*

We employed supervised **Machine Learning** (ML) techniques to estimate the vertical acceleration (in place of gravity variations, see next paragraph on Training Platform) from the outputs of the multi-sensor system. We formulated the problem as a regression task, in which the desired output was provided by the vertical acceleration derived from the reference sensors of the training platform.

We investigated several supervised regression algorithms. We adopted Linear Regression and its regularized variants (Ridge and Lasso) as baseline models, while nonlinear methods—including Support Vector Regression and ensemble-based regressors—were considered to capture more complex relationships between sensor outputs and vertical acceleration. We trained all ML models using synchronized measurements from reference sensors, accelerometer and gyroscope, as well as additional features derived from the latter, such as time derivatives of the gyroscope signals and nonlinear terms (e.g., squared components). We performed ML model training and hyperparameter tuning using a cross-validation strategy to enhance generalization performance and mitigate overfitting.

For comparison purposes, we computed the vertical acceleration using an analytical model (System model) based on the integration of angular rates to estimate platform orientation and on a parametric reconstruction of the gravity projection along the vertical axis.

The **Training Platform** for this experiment adopted the full configuration previously introduced in Figure 4 of Section 4.1.3. A key challenge in this setup was how to generate and measure a time-varying reference signal representative of gravity variations and use it as the training label for supervised ML. In principle, one option would have been to perform a field survey over an area whose gravity field had already been mapped before, so that reference labels were provided. However, this approach would have introduced major practical limitations, including the impossibility of independently controlling the relevant disturbances. A second option, achievable in a laboratory, would have been to physically vary gravity by moving large masses. Yet this approach was also impractical. To produce a gravity variation of only 1 mGal ($\sim 10^{-5}$ m/s$^2$), a lead sphere about 3 m in radius and weighing roughly 1500 tons would be needed. The sphere would have to be moved back and forth by about 10 m, from nearly touching the gravimeter to farther away, to reduce the signal to about 0.1 mGal ($\sim 10^{-6}$ m/s$^2$).

To avoid such logistical complexity, we leveraged Einstein's equivalence principle, which states the equivalence between gravity and apparent accelerations. Based on this principle, we designed the platform so that it could oscillate vertically, thereby generating controlled apparent accelerations. We calculated that a sinusoidal oscillation, with an amplitude of 0.1 m and a frequency of 0.05 Hz, produces a maximum acceleration of approximately 1000 mGal ($\approx 0.01$ m/s$^2$). This value should be interpreted as an illustrative order-of-magnitude estimate based on motion parameters compatible with our platform, not as a requirement or as the amplitude used in all experimental tests. This approach therefore enabled highly flexible and reproducible variations of the measurand. It is important to clarify, however, that this experiment only tested the first stage of airborne gravimetry: in a real application, a second stage relying on GNSS systems is required to separate gravity from the apparent accelerations.

The mechanical structure of the training platform consisted of an aluminum plate, on which the multi-sensor system was mounted (Figure 8 and Figure 9). Three linear actuators (SMC LDZBB3L-100A3) positioned at the vertices of an isosceles right triangle supported the plate. The tips of the actuators acted as the feet of the platform, each capable of a vertical stroke of about 0.1 m. This configuration enabled three degrees of freedom and two modes of motion. 1) Vertical (altitude) oscillations, required to simulate gravity variations, induced by driving all actuators together. 2) Pitch and roll tilts of the plate induced by differentially driving the actuators.

We powered the actuators through a programmable bench power supply (Rohde & Schwarz NGE100). Their length was regulated by a software-based PID control loop running on a dedicated Raspberry Pi. For feedback, the system relied on three ultrasonic distance meters (HC−SR04), one per actuator, which continuously measured the length of each foot. This closed-loop control regulated the modulation of both altitude and tilt, enabling the platform to reproduce the programmed vertical oscillations and rotations.

To implement the reference sensor—essential for providing labeled data in supervised machine learning—the vertical displacement of the platform was also monitored with high precision. To this end, we equipped each actuator tip with a Linear Encoder (LE) - Gefran PZ−34−A−125 potentiometer-. Their outputs were digitized using a high-resolution ADC (Texas Instruments ADS1263) and acquired by an ESP32 microcontroller. While moving, the amplitude spectral density of the background noise of each encoder was about 0.2 mm/√Hz. By combining these measurements with the known geometry of the platform, we could reconstruct the altitude of any selected point, such as the location of the gravity sensor, (Iafolla et al., 2024a). Once the altitude time-series was obtained, the vertical acceleration was computed by applying a second-order numerical differentiation based on a Savitzky–Golay filter. This filter simultaneously smooths the data and estimates the second time derivative, thereby reducing the amplification of high-frequency noise. While moving, the amplitude spectral density of the background noise of the estimated vertical acceleration was about $2\times10^{-4}$ m/s$^2$/√Hz. This acceleration served as the independent reference signal, ensuring that the training datasets were properly labeled for machine learning.

The sensors and actuators were selected according to the functional requirements of this preliminary laboratory implementation. The actuators had to generate repeatable vertical displacements and controlled pitch/roll rotations within the safe operating range of the mechanical setup. The reference sensors had to measure the platform displacement with sufficient resolution to derive the vertical acceleration labels, and all signals had to be synchronized with the multi-sensor system. The purpose of this setup was therefore to implement and test the proposed framework under controlled and sufficiently challenging laboratory conditions, rather than to reproduce a complete airborne gravimetry calibration facility.

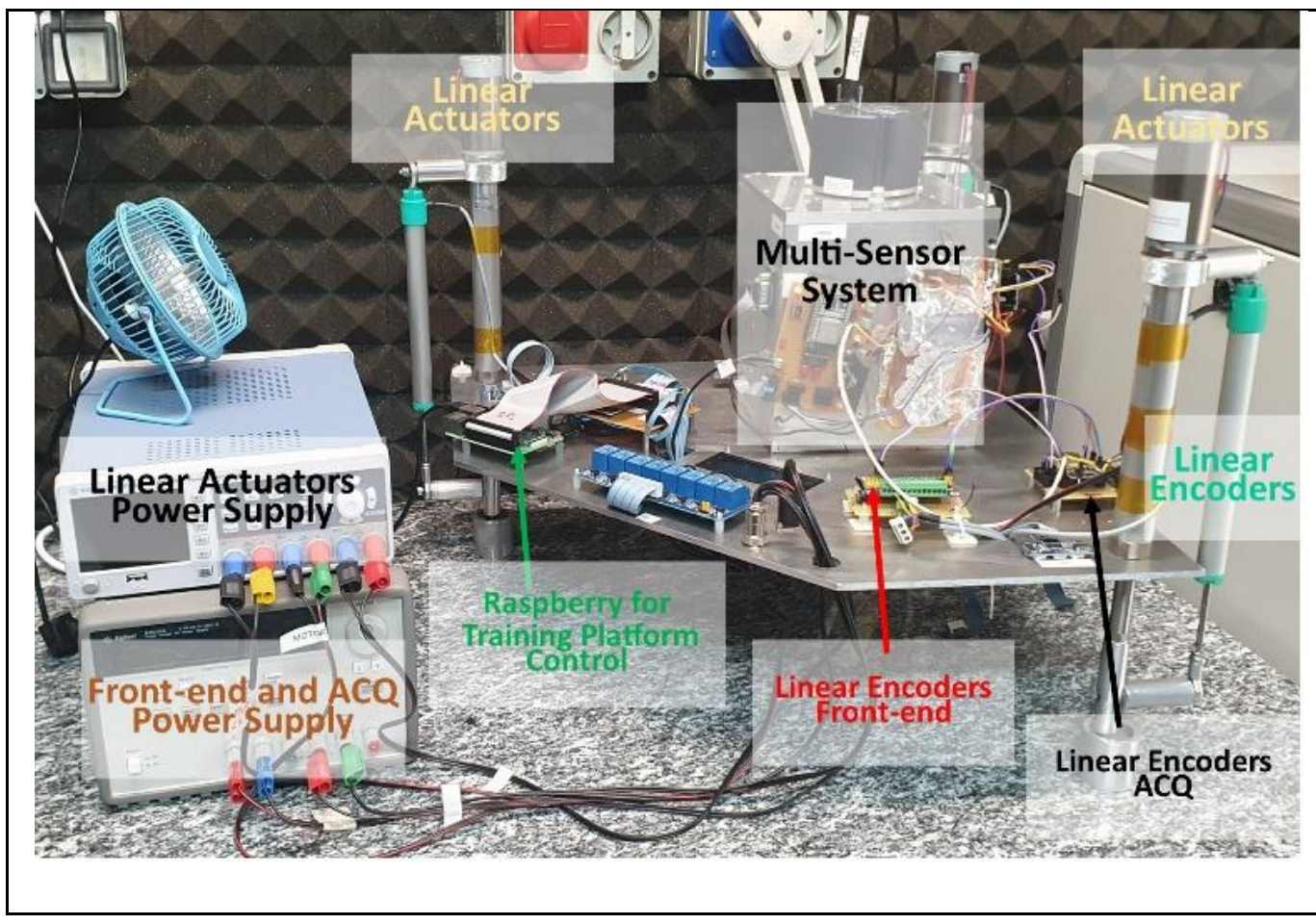

*Figure 8. Photo of the side view of the training platform used for the - Pitch and roll rejection for vertical acceleration measurement – experimental setup.*

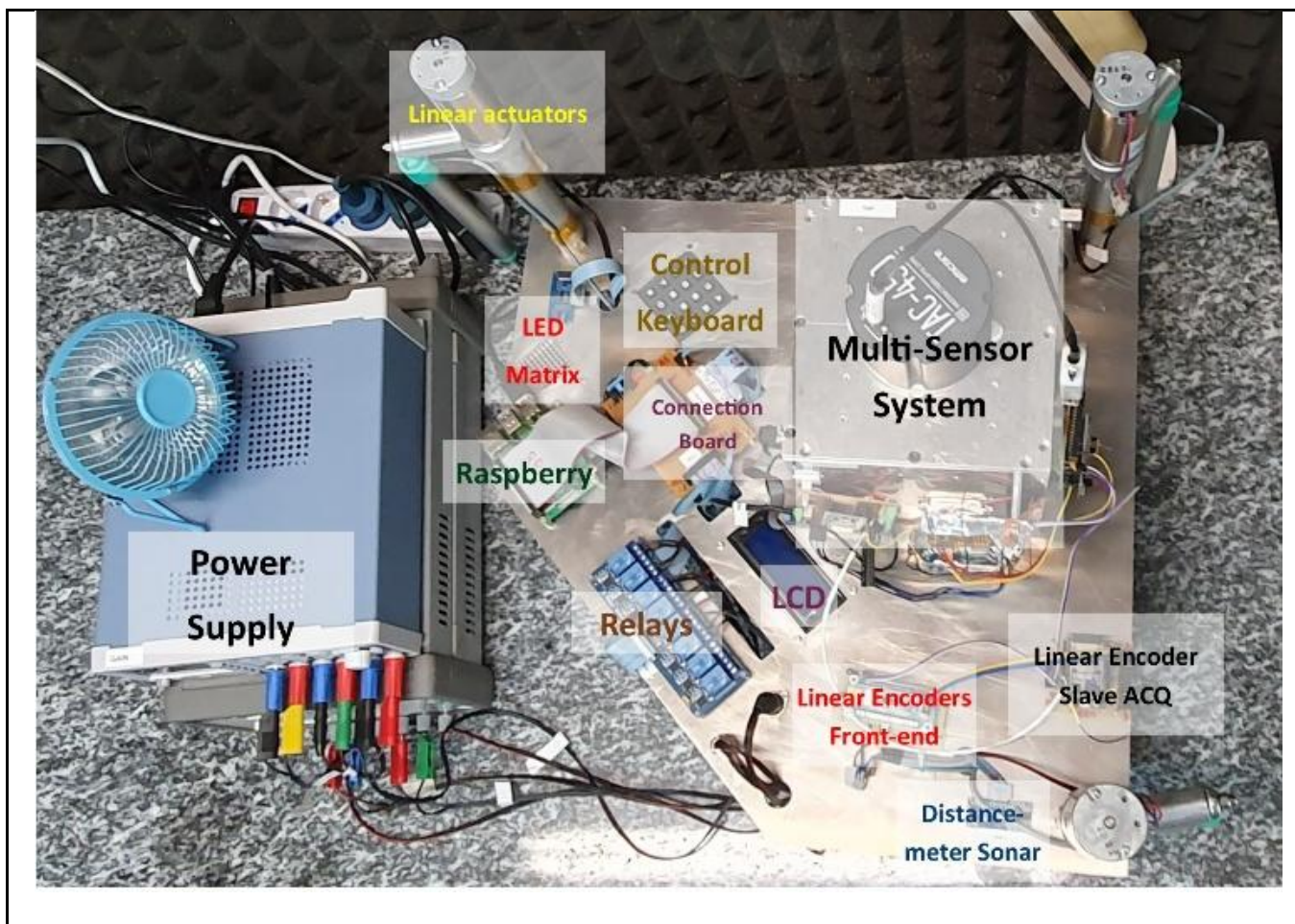

*Figure 9. Photo of the top view of the training platform used for the - Pitch and roll rejection for vertical acceleration measurement – experimental setup.*

# 5 Results

This section presents the results of two experimental activities related to the applications described in Section 3.

The temperature-rejection experiment (see Section 3.1) is a previously validated case study revised in this paper to illustrate the application of the proposed framework to thermal and thermal-gradient disturbances. The present discussion focuses on the role of this experiment within the general compensation framework; experimental data and an extensive performance analysis have been reported previously (Iafolla et al., 2024c, 2023).

The pitch/roll rejection experiment (see Section 3.2) is an original case study demonstrating the extension of the proposed framework to rotation-induced disturbances and vertical acceleration reconstruction. The results reported below are used to prove the principle, to assess the feasibility of the approach, and to identify the main requirements for future implementations based on upgraded experimental platforms. The corresponding experimental dataset is publicly available (Iafolla et al., 2024a).

## 5.1 Results of temperature rejection experiment

As illustrated by the time-series examples reported in Figure 10, the compensation signals (blue traces) closely followed the target output (black traces) for both the gravity-aligned component (z) and a horizontal component (x), with residual fluctuations (red traces) markedly reduced over the shown time interval. Quantitative comparisons with conventional single-thermometer compensation are reported in Table 1 for component x and in **Errore. L'origine riferimento non è stata trovata.** for component z. These comparisons were obtained using the same experimental dataset; the single-thermometer case was implemented as a data processing baseline, not as a separate hardware setup. When only the thermometer placed on the x sensor was used, linear regression yielded an RMSE of $6.6\times10^{-6}$ m/s$^2$ and $R^2 = -0.006$. In contrast, the multi-thermometer approaches reduced the RMSE to approximately $2.0–2.1 \times 10^{-6}$ m/s$^2$ and increased $R^2$ to approximately 0.90–0.91. Results for vertical component z (Table **2**) show stronger dependence to temperature of the sensor ($R^2 = 0.947$), anyway, the multi-thermometer approaches improved both RMSE and $R^2$.

| Compensation approach | Input thermometers | RMSE [m/s$^2$] | $R^2$ |
|---|---|---|---|
| Linear regression, thermometer on x sensor | 1 | $6.6 \times 10^{-6}$ | −0.006 |
| Multivariate linear regression | 11 | $2.1 \times 10^{-6}$ | 0.900 |
| FFNN, multi-thermometer input | 11 | $2.0 \times 10^{-6}$ | 0.910 |

Table 1. Compensation results for component x of the accelerometer

| Compensation approach | Input thermometers | RMSE [m/s$^2$] | $R^2$ |
|---|---|---|---|
| Linear regression, thermometer on z sensor | 1 | $1.5 \times 10^{-5}$ | 0.947 |
| Multivariate linear regression | 11 | $2.7 \times 10^{-6}$ | 0.998 |
| FFNN, multi-thermometer input | 11 | $2.5 \times 10^{-6}$ | 0.999 |

Table 2. Compensation results for component z of the accelerometer

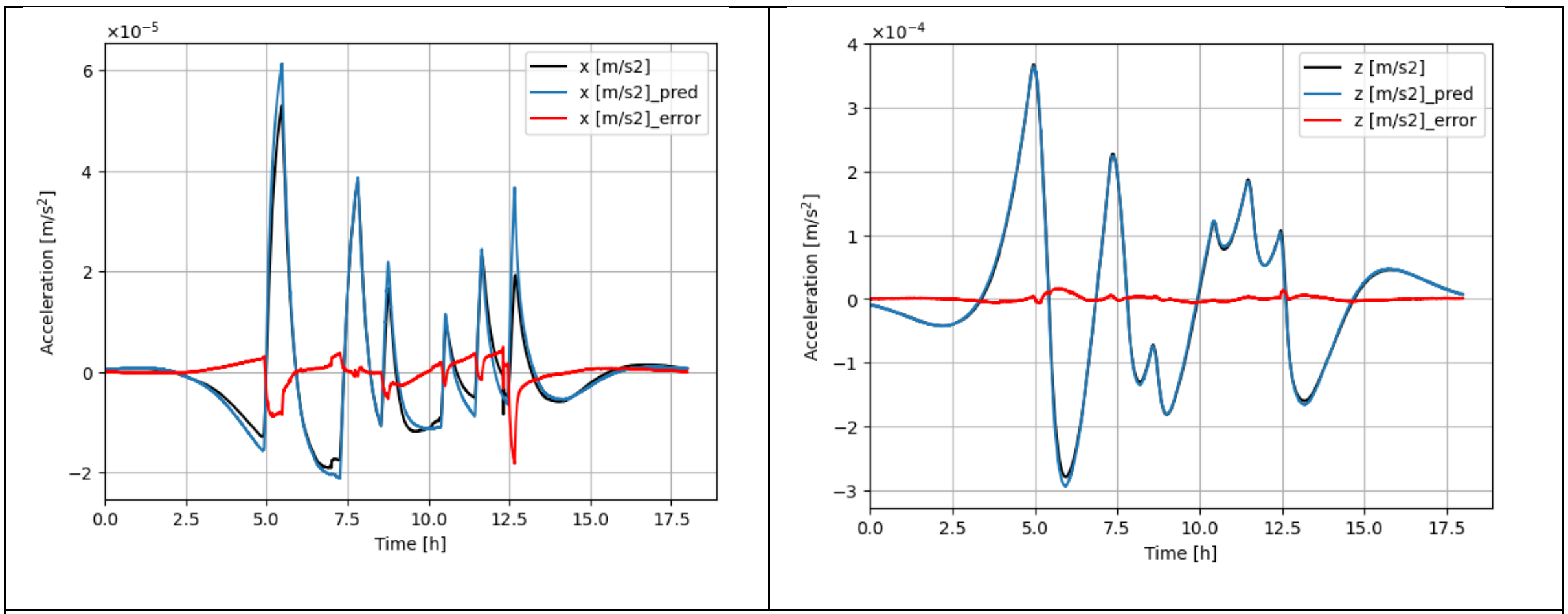

*Figure 10. Left: Time plots of desired output (x [m/s2]), predicted output (x [m/s2]_pred) and their difference (i.e., the error, x [m/s2]_error). Right: Time plots of desired output (z [m/s2]), predicted output (z [m/s2]_pred) and their difference (i.e., the error, z [m/s2]_error).*

These results indicate that using multiple temperature measurements distributed across the enclosure was essential to capture thermal gradients, which contributed substantially to the observed output variations, especially on the horizontal axes.

## 5.2 Results of pitch and roll rejection for vertical acceleration measurement

We emphasize that this experiment is a proof of principle for vertical acceleration reconstruction under controlled pitch/roll disturbances, not a complete demonstration of airborne gravimetry.

Table 3 summarizes the main experimental parameters and reconstruction-performance metrics over the full dataset. The reference vertical acceleration had a peak-to-peak amplitude of approximately 900 mGal ($9{\cdot}10^{-3}$ m/s$^2$), whereas the accelerometer z-axis output had a peak-to-peak amplitude of approximately 10 Gal ($100{\cdot}10^{-3}$ m/s$^2$). The latter signal was therefore more than 10 times larger than the reference vertical acceleration, highlighting the difficulty of the reconstruction problem. This occurred because the accelerometer's z-axis output included both the vertical acceleration of the platform and the much larger contribution produced by the rotation-induced projection of gravity onto the sensor axis. In other words, when pitch and roll rotations were applied, the accelerometer z-axis did not coincide with the vertical acceleration in the laboratory frame.

Figure 11 reports representative time series obtained during the pitch and roll rejection experiment. The red trace of the upper subplot shows the output of the accelerometer's z-axis. In the same subplot, the black trace was obtained using an analytical system model. As first step, this model reconstructs the projection of the gravity vector onto the accelerometer's z-axis (which coincides with the training-platform z-axis) from gyroscope-derived orientation angles. Thus, the black trace can be interpreted as the disturbance affecting the accelerometer when its sensitive axis is continuously misaligned with respect to the local vertical (i.e., the gravity vector) due to platform rotations. In this test, the peak-to-peak amplitude of this contribution was approximately $15{\cdot}10^{-3}$ m/s$^2$. The difference between the accelerometer output (red trace) and the reconstructed gravity projection (black trace) corresponds to the component of the apparent (inertial) acceleration along the accelerometer's z-axis. Within the analytical system model, this quantity can be then projected into the laboratory reference frame—using the same gyroscope-derived orientation angles—to obtain an estimate of the vertical acceleration of the training platform.

| Metric | Value | Description |
| --- | --- | --- |
| **Reference vertical acceleration**, 0.03–0.3 Hz band | peak-to-peak ≈ $9{\cdot}10^{-3}$ m/s$^2$; std = $1{\cdot}10^{-3}$ m/s$^2$ | Amplitude range of the target signal used as ML label |
| **Accelerometer z-axis output**, 0.03–0.3 Hz band | peak-to-peak ≈ $100{\cdot}10^{-3}$ m/s$^2$; std = $6{\cdot}10^{-3}$ m/s$^2$ | Maximum measured signal including vertical acceleration and rotation-induced gravity projection |
| **Pitch/roll angular range**, <0.03 Hz | ±5.7 deg; std = 1.5 deg | Overall range of imposed rotational disturbances |
| **Pitch/roll angular range**, 0.03–0.3 Hz band | ±2.9 deg; std = 0.9 deg | Dynamic rotational disturbance range in the frequency band most relevant to the experiment |
| **Dataset duration** | 5.6 h | Amount of data available for training and testing |
| **Analytical system-model RMSE** | $2{\cdot}10^{-3}$ m/s$^2$ | Baseline reconstruction error |
| **ML prediction RMSE** | $0.9{\cdot}10^{-3}$ m/s$^2$ | Reconstruction error after ML compensation |
| **RMSE reduction** | 51% | Quantitative improvement relative to the analytical model |

Table 3. Summary of the main experimental parameters and reconstruction performance for the pitch/roll rejection experiment over the full dataset.

The middle subplot of Figure 11 shows the vertical acceleration expressed in the laboratory reference frame. As discussed in Section 4.2.2, this signal represents the proxy for gravity variations adopted in the training platform. Three traces are reported. The blue trace corresponds to the reference vertical acceleration derived from the linear encoders and exhibits a peak-to-peak amplitude of about $2{\cdot}10^{-3}$ m/s$^2$, which is approximately 7.5 times smaller than the disturbance shown in the upper subplot, resulting in a signal-to-disturbance ratio of about 0.13. The orange trace (“System model”) represents the vertical acceleration estimated using the analytical system model. The green trace corresponds to the vertical acceleration estimated by the machine learning algorithm fed with the multi-sensor outputs. The bottom subplot of Figure 11 shows the discrepancies between reference vertical acceleration and the other two respectively.

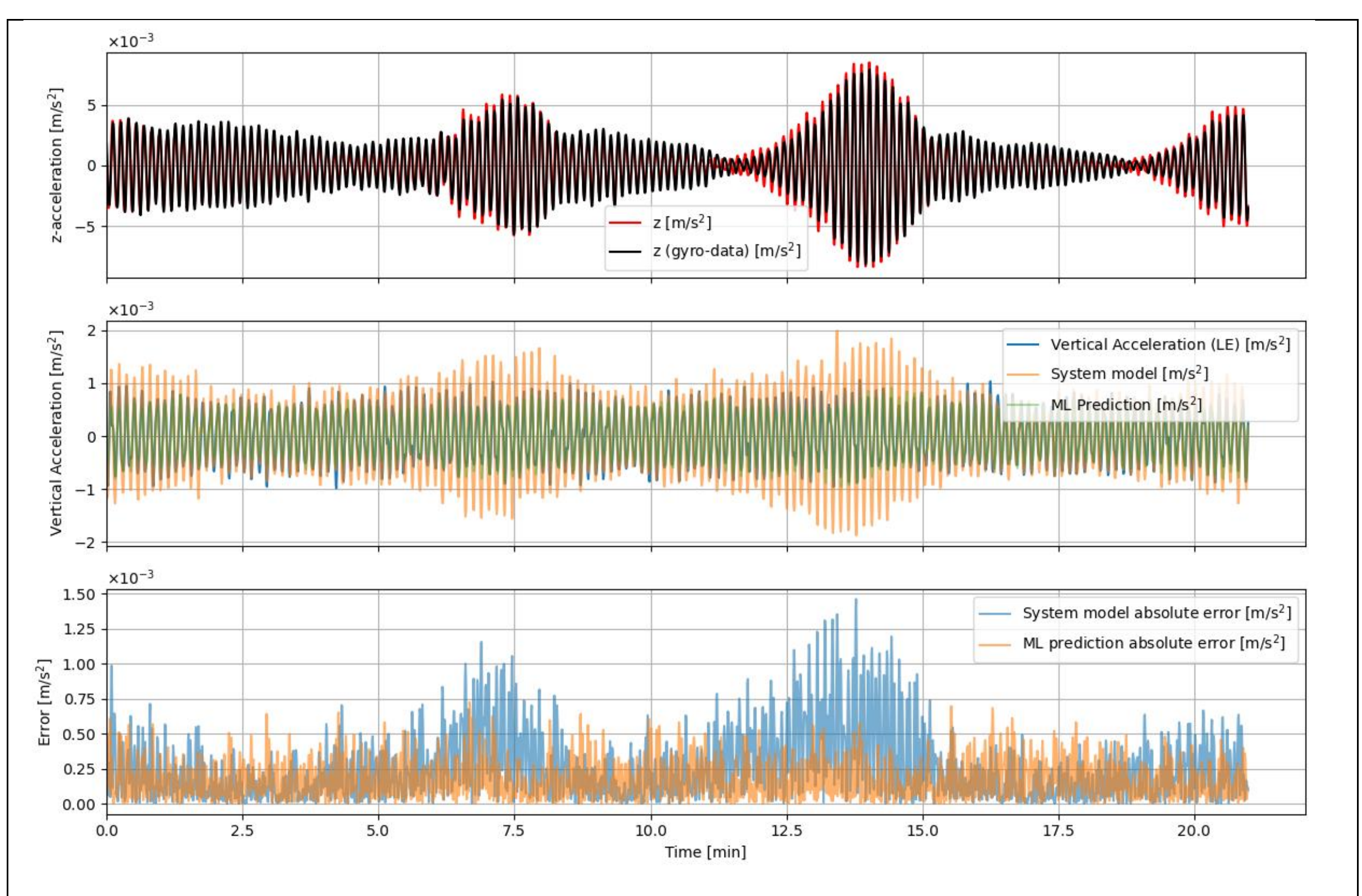

*Figure 11. Time series of z-axis and vertical acceleration. Top: z-axis (training platform reference frame) acceleration measured by the accelerometer (red trace) and reconstructed using an analytical system model based on gyroscope-derived orientation angles (black trace). Middle: vertical acceleration in the laboratory reference frame. The blue trace represents the reference signal derived from linear encoders (LE); the orange trace corresponds to the analytical system model; the green trace shows the prediction obtained with the ML algorithm fed with multi-sensor outputs. Bottom: discrepancies between the reference signal and the model-derived signals.*

Figure 12 presents a scatter plot comparing the vertical acceleration predicted by the analytical system model (orange dots) and by the ML approach (blue dots) against the reference values obtained from the linear encoders. In the ideal case of perfect reconstruction, all points would lie on the dashed black line (y = x). Deviations from this line indicate reduced accuracy.

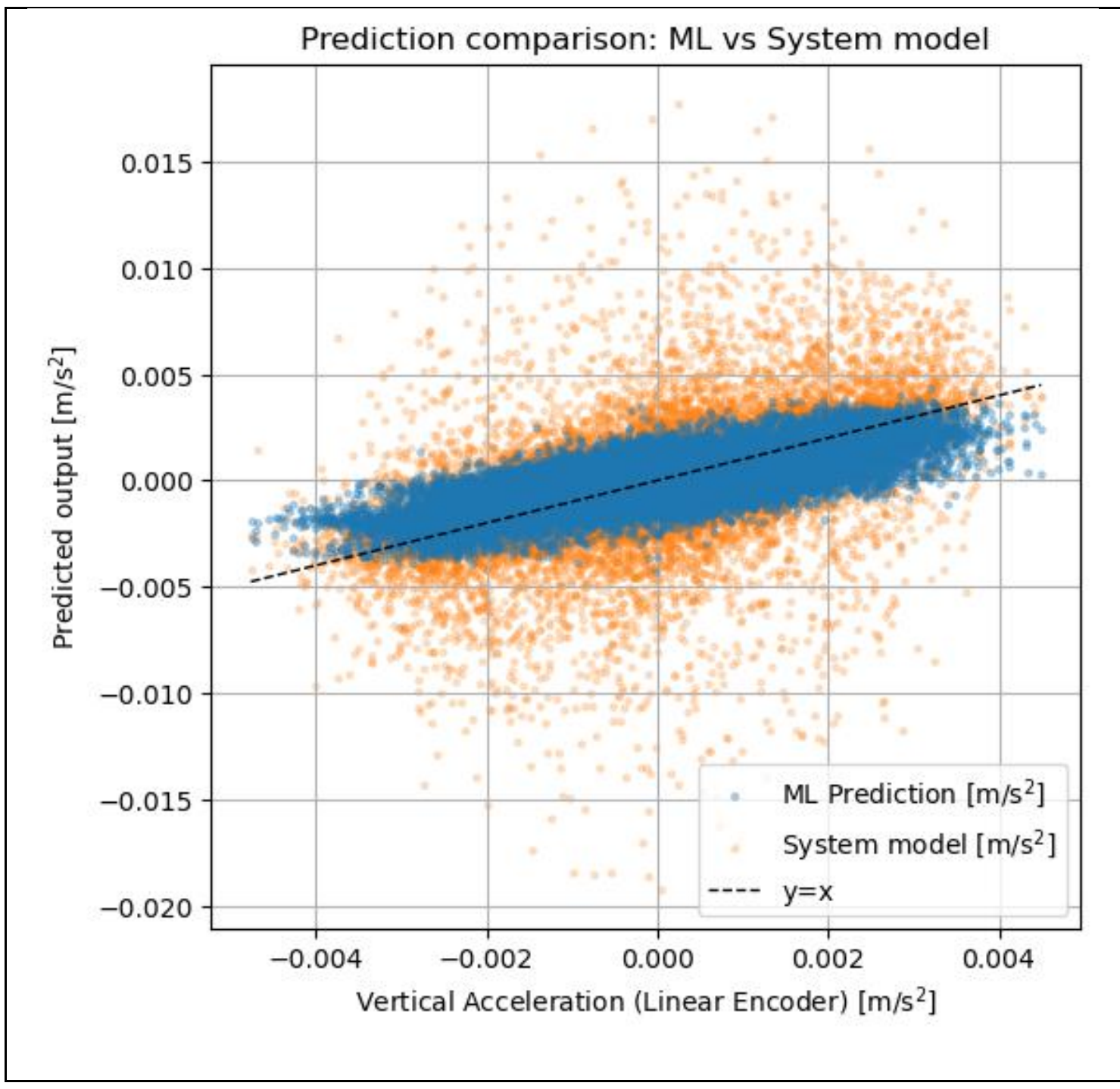

*Figure 12. Scatter plot of predicted versus reference vertical acceleration. Orange dots correspond to predictions obtained using the analytical system model, while blue dots correspond to ML-based predictions. The dashed black line represents the ideal case (y = x). The ML model shows a substantially lower dispersion than the System Model, although its point cloud is slightly tilted (non-unitary gain) with respect to the ideal line because the model is optimized to minimize prediction error rather than to enforce a unitary gain.*

Finally, Figure 13 shows the distributions of the estimation errors, defined as the difference between the vertical acceleration predicted by either the analytical system model or the ML approach and the corresponding reference value. Error distributions with wider spread or significant displacement from zero indicate poorer performance. The comparison highlights the improved accuracy achieved by the ML-based reconstruction under the tested conditions.

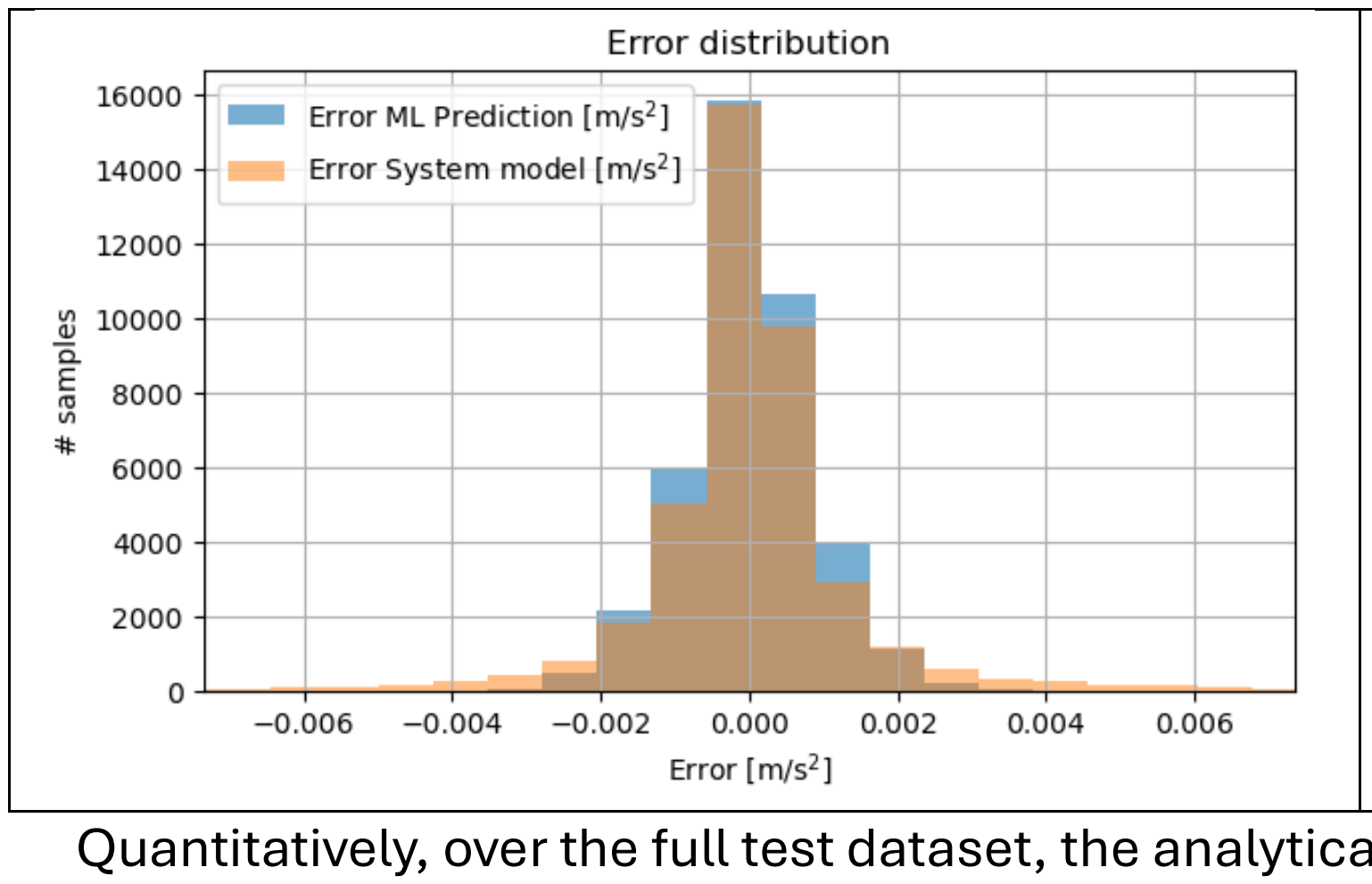

*Figure 13. Histograms of the error distributions of the vertical acceleration estimates with respect to the reference values. Errors obtained using the analytical system model are shown in orange, while those obtained using the ML approach are shown in blue.*

Quantitatively, over the full test dataset, the analytical system model yielded an RMSE of 200 mGal ($2{\cdot}10^{-3}$ m/s$^2$), whereas the ML-based reconstruction yielded an RMSE of 90.0 mGal ($0.9{\cdot}10^{-3}$ m/s$^2$). This corresponds to an RMSE reduction of ~55% indicating that, under the tested laboratory conditions, the ML model improved the reconstruction of the reference vertical acceleration with respect to the analytical baseline.

# 6 Discussion

This study addressed a critical gap in high-accuracy gravity measurement systems: the difficulty of compensating for complex, interacting disturbances in demanding applications such as airborne gravimetry. Conventional approaches—including inherent insensitivity, opposing input, analytical corrections, or physical insulation—offer valuable tools but face limitations when confronted with nonlinear effects, simultaneous presence of multiple disturbances, hardware imperfections, or practical constraints such as space and power requirements for passive or active insulation systems. The presented method was conceived as an additional tool to overcome these challenges by combining multi-sensor acquisition, machine learning, and a dedicated training platform.

The experiments conducted in this study supported the feasibility of the proposed approach across different classes of applications and disturbances. In the temperature rejection experiment, machine learning successfully modeled complex sensor responses using multidimensional inputs (three accelerometric channels and eleven thermometer readings). In the pitch and roll rejection experiment, machine learning effectively accounted for disturbances induced by reference-frame rotations through gyroscope measurements. Beyond demonstrating feasibility, these experiments highlighted that the training platform enables highly versatile data generation. For example, by deliberately imposing thermal gradients or controlled tilts of the platform, we could generate training and testing datasets that mirrored operational conditions while spanning the full expected range of variability. This ability to design data distributions in the laboratory represents a fundamental advantage over field-based data collection, where disturbance variability is neither predictable nor controllable.

Compared with conventional approaches, our method provides two main advantages. First, it automatically considers system-specific characteristics—such as sensor nonlinearities, misalignments, assembly-related imperfections, or other factors that would otherwise remain unmodeled—within the compensation model. Second, the training platform ensures that datasets cover the full range of expected conditions, thereby reducing the risk of extrapolation that would arise if training data were collected directly in the field (e.g., when temperature ranges during a measurement campaign are too narrow). These features position our approach as a complementary tool to conventional methods, extending disturbance rejection capabilities to regimes where analytical modeling and physical compensation or insulation become impractical. For example, in the pitch and roll rejection experiment, the improved agreement between the ML-based reconstruction and the reference signal, compared with the analytical system model, suggests that the data-driven ML model captured residual system-specific effects not explicitly included in the analytical formulation.

The reconstruction errors reported in Section 5 reflect not only ML prediction accuracy, but also errors introduced by the multi-sensor system and the training platform. In the temperature-rejection experiment, the dominant contribution was likely the incomplete sampling of the temperature field, where accelerometer and thermometer noise, acquisition-electronics drift, synchronization errors, and nonthermal environmental signals may have provided additional contributions. In the pitch/roll experiment, the main limitation was likely the training platform, particularly the limited accuracy of the linear encoders and uncertainties in the platform geometry. Additional contributions may have arisen from sensor misalignment, poor time-resolution of the gyroscope, synchronization errors, and numerical differentiation. A complete uncertainty budget would require a separate characterization of these terms. Importantly, the availability of labeled test data acquired under controlled conditions is a key advantage of our framework because it allows the error to be evaluated as a function of the imposed operating regime, such as the amplitude and rate of temperature variations or the pitch and roll ranges. This makes it possible to identify the conditions under which the overall measurement and compensation system remains reliable and those under which its error increases.

Nonetheless, applying our method is not a plug-and-play procedure, as each application requires *ad hoc* development of both hardware and training strategies. This also applies to the selection of the ML algorithm: the proposed framework does not prescribe a specific category of models a priori. For instance, the choice between conventional regression algorithms and artificial neural networks must be made according to the specific measurement problem, such as: the complexity of the disturbance–response relationship; the available training data; and the desired balance between predictive performance, robustness, and interpretability. An example of *ad hoc* strategy was shown for the pitch and roll rejection experiment. A key challenge was how to vary the measurand (gravity) in the laboratory. This was addressed by exploiting the equivalence principle between gravity and apparent accelerations, designing a platform capable of vertical oscillations. In the temperature rejection experiment, a tailored strategy was implemented by equipping the accelerometer with multiple thermometers placed to capture thermal gradients within the sensor enclosure. These examples illustrate how specific experimental solutions were developed to meet the requirements of each application.

# 7 Conclusions

This work advanced the understanding of how high-accuracy gravity and acceleration measurements can be achieved in environments characterized by strong and interacting disturbances. Our method—based on a multi-sensor system, machine learning, and a dedicated training platform—was conceived as a general framework for disturbance compensation and was demonstrated in two experimental contexts: temperature rejection in static accelerometers and pitch/roll rejection for vertical acceleration measurements.

The experiments confirmed that, while the method provides general guidelines, each application requires dedicated solutions. Our examples highlight that an experimental phase is always necessary to define both hardware configurations and training strategies.

Both case studies showed that the proposed framework is feasible, but their main contribution was methodological rather than operational. The experiments were intended to define and test the practical procedures required to implement the framework, rather than to demonstrate a field-ready airborne or seaborne gravimeter. This was particularly evident in the pitch/roll experiment, where a key achievement was the development of a laboratory strategy for generating a controlled, time-varying reference signal representative of gravity variations. Under these conditions, the ML-based reconstruction reduced the RMSE from 200 mGal ($2 \cdot 10^{-3}$ m/s$^2$), obtained with the analytical system model, to 90 mGal ($0.9 \cdot 10^{-3}$ m/s$^2$). Similarly, temperature compensation experiment demonstrated that our method could reduce the temperature induced error on x axis from $6.6 \cdot 10^{-6}$ m/s$^2$ to $2.0 \cdot 10^{-6}$ m/s$^2$. These results were encouraging for demonstrating feasibility and for understanding how the method should be applied. However, they should not be interpreted as general performance specifications of the framework or as a measure of operational airborne-gravimetry accuracy.

Future experiments should therefore proceed stepwise from improved laboratory validation to increasingly realistic dynamic tests. More advanced training platforms, such as commercial hexapods, could provide higher accuracy and more flexible actuation of rotations and displacements. Sensor integration should also be refined, for example by co-locating gyroscopes and accelerometers in a common IMU structure or by adding thermometers on critical electronic components, such as preamplifiers and reference voltage generators. Field measurements in air or at sea represent a natural subsequent step, but they should follow a more complete validation of the training platform, sensor integration, synchronization architecture, and data-processing chain.

A further refinement of the machine learning pillar could be obtained with *model-based machine learning*. In this approach, analytical or physics-based models are embedded into the learning process, constraining predictions while leaving the ML algorithm to capture residual effects such as hardware imperfections, sensor nonlinearities, or other system-specific features not foreseen during analytical modelling. For example, in multi-sensor gravimetry the known dynamical relations between accelerometers, gyroscopes, and the reference frame can be explicitly incorporated. By combining physical modeling with data-driven components, this hybrid strategy constrains extrapolation through established physical principles and enhances robustness under conditions not fully represented in the training data.

In conclusion, our method should be regarded as a broadly applicable framework rather than a ready-made solution. Its generality allows extension well beyond the case studies presented here, but its implementation must always be tailored to the specific measurement context. Future work should therefore focus on developing optimized training platforms, refining sensor integration, and exploring hybrid model-based machine learning approaches, ultimately laying the foundation for more accurate and reliable measurement systems in disturbance-rich environments.

# Declarations

**Experimental Data**

Experimental data are publicly available here: https://doi.org/10.17632/crvy9jgb6g.1

**Declaration of generative AI and AI-assisted technologies in the writing process**

During the preparation of this work the authors used ChatGPT of OpenAI in order to improve language and readability of the text. After using this tool/service, the authors reviewed and edited the content as needed and take full responsibility for the content of the publication.

**Conflict of Interest statement**

Lorenzo Iafolla, Francesco Santoli, Roberto Carluccio, Stefano Chiappini, Emiliano Fiorenza, Marco Lucente, and Massimo Chiappini have patent issued to WO2025052255 - MULTI-SENSOR SYSTEM FOR MEASURING ACCELERATION. Other authors declare that they have no known competing financial interests or personal relationships that could have appeared to influence the work reported in this paper.

# Funding

This research received partial funding from the “Gravimetro Aereo INtelligente” (GAIN) project, supported by Regione Lazio (Italy) through the European Regional Development Fund (POR FESR Lazio 2014–2020), and from the SAKURA research line, under the ROSE infrastructural project of INGV (OB.FU.: 1215.010), funded by the Italian Ministry of University and Research.